\documentclass[aip,rsi,reprint]{revtex4-1} 

\usepackage[english]{babel}
\usepackage[pdftex]{graphicx}

\usepackage{graphicx}

\begin{document}


\title{High efficiency positron accumulation for high-precision magnetic moment experiments} 



\author{S. Fogwell Hoogerheide}
\altaffiliation{current address: National Institute of Standards and Technology, Gaithersburg, MD 20899}

\author{J. C. Dorr}
\altaffiliation{current address: Honeywell Automation and Control Solutions, Golden Valley, MN 55422}

\author{E. Novitski}

\author{G. Gabrielse}
\affiliation{Harvard University, Cambridge, Massachusetts, 02138, USA}


\date{\today}

\begin{abstract} 
Positrons are accumulated within a Penning trap designed to make more precise measurements of the positron and electron magnetic moments. The retractable radioactive source used is weak enough to require no license for handling radioactive material and the radiation dosage one meter from the source gives an exposure several times smaller than the average radiation dose on the earth's surface. The 100 mK trap is mechanically aligned with the 4.2 K superconducting solenoid that produces a 6 tesla magnetic trapping field with a direct mechanical coupling.   
\end{abstract}


\maketitle 


\section{Introduction}
A great triumph of the standard model of particle physics \cite{StandardModelTriumph} is the remarkable, part-per-trillion agreement between the most precise measurement of a property of an elementary particle and the standard model's most precise prediction. The electron magnetic moment is measured to 3 parts in $10^{13}$ using one electron in a 100 mK Penning trap  \cite{HarvardMagneticMoment2008} .   The standard model's prediction of the same quantity comes from the Dirac equation, tenth order quantum electrodynamics (QED) \cite{TenthOrderQED2012}, hadronic contributions \cite{MuonLightByLight}, and weak contributions \cite{Weak1972,Marciano:1996,Weak2002,Weak2003} as well as a measured value of the fine structure constant \cite{RbAlpha2011}.  

The positron magnetic moment could be measured just as precisely with a positron suspended in the same trap apparatus. Comparing the positron and electron moments would test the CPT theorem (charge conjugation, parity and time reversal) with leptons about 15 times more precisely than the best previous comparison of a lepton and antilepton \cite{DehmeltMagneticMoment}. This fundamental theorem of the standard model predicts that the positron moment should be opposite in sign but equal in magnitude to the electron moment.  

In some ways, positrons from a radioactive source are compatible with the cryogenic, high vacuum environment needed for the most precise measurements.  A Na$^{22}$ source has a 2.6 year half-life that is long enough to make measurements but short enough to avoid long-term disposal issues.  The activity of the Na$^{22}$ sources used previously  to trap positrons range from about 0.5 mCi to 75 mCi.  The 0.5 mCi source, the smallest of these, was fixed to a 4.2 K trap to accumulate trapped positrons at a rate of 0.8 positron/min/mCi \cite{UwPositrons} -- just high enough for measurements that require only a single trapped positron.  

One of two drawbacks, however, is that even this source activity is fifty times higher than the 10 $\mu$Ci safety threshold that necessitates a license and a higher level of special handling for the radioactive materials \cite{Na22LicenseThreshold} -- a significant complication for doing the measurement.  The second drawback, not well understood, is that a source fixed near the 4.2 K electrodes was observed to result in spontaneous electron loading during electron measurements.   

\begin{figure}
 \includegraphics[width=3in]{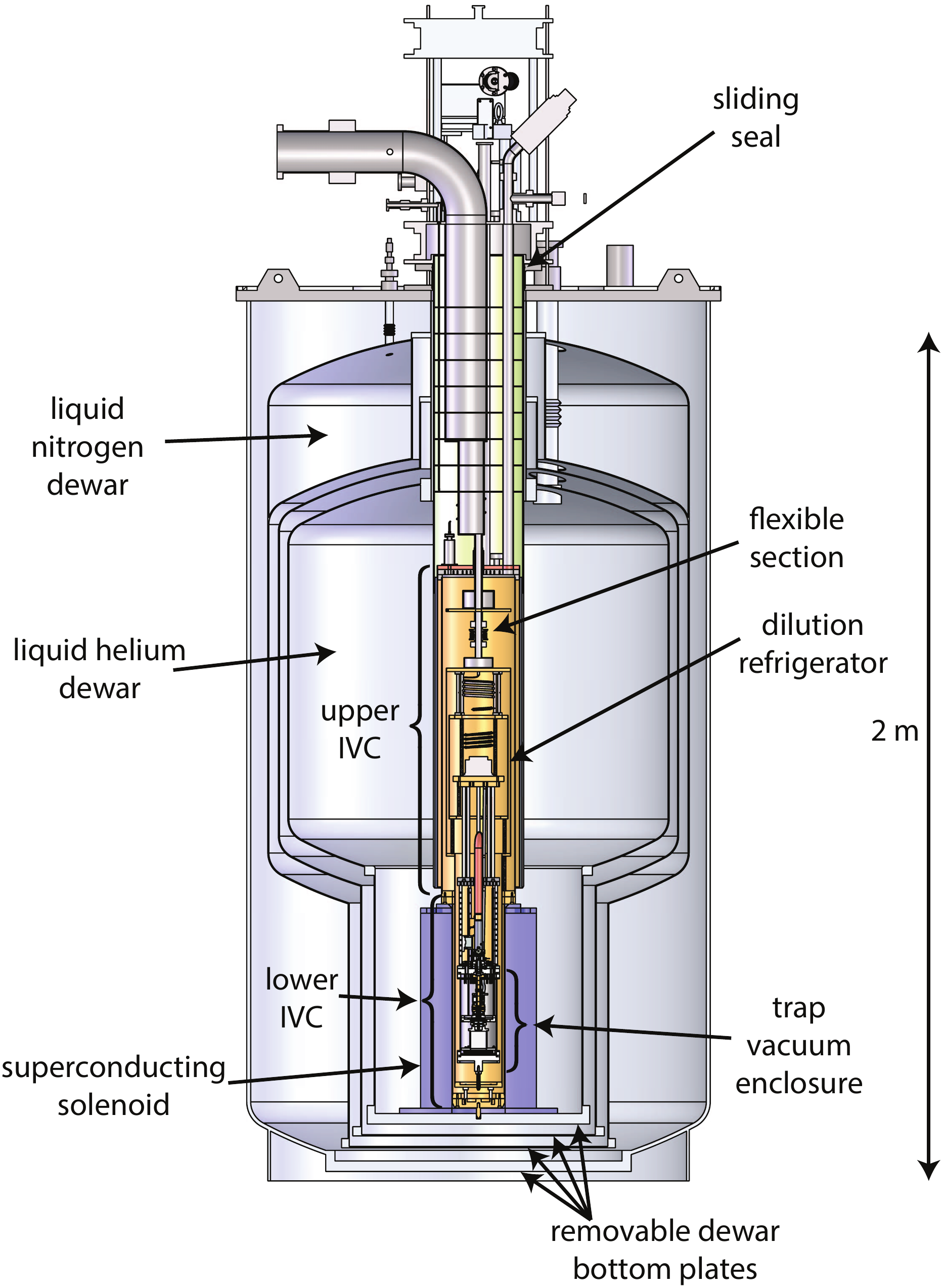}
 \caption{Apparatus designed for improved comparisons of the magnetic moments and charge-to-mass ratios of the electron and positron.}\label{fig:WholeApparatus}
 \end{figure}

Here we report using a 6.5 $\mu$Ci source (about 77 times weaker) to load positrons at a rate above 150 positrons/min/mCi (about 190 times higher) into a 100 mK trap within a new apparatus (Fig.~\ref{fig:WholeApparatus}).   The loading mechanism used was initially developed for efficiently accumulating positrons directly into high vacuum for antihydrogen production \cite{PositronsFromPositronium} from very large radioactive sources. The accumulation rate is higher than needed for precise measurements with a trapped positron, so useful positron loading should continue for many Na$^{22}$ half-lives.  

In fact, the 6.5 $\mu$Ci source is below the 10 $\mu$Ci threshold that triggers the mentioned licensing requirement\cite{Na22LicenseThreshold}.  No cumbersome heavy metal shielding is required since a point source of this activity at a distance of 1 meter gives a does rate of 0.01 mrem/hour (0.1 $\mu$Sv/hour) -- an annual dose of 90 mrem (900 $\mu$Sv). This is much smaller than the average natural background rate of 310 mrem (3.1 mSv) from natural sources and the 310 mrem (3.1 mSv) from manmade sources \cite{NaturalBackground}.  It is also lower than the 100 mrem (1 mSv) that the Nuclear Regulatory Commissions (NRC) has established as the maximum yearly radiation dose to which one of its licensees can expose an individual member of the general public\cite{NaturalBackground}. 

Spontaneous electron loading is prevented by making it possible to withdraw the radioactive source away from the trap.  In additions, the positron loading is carried out with trap electrodes that are kept at 100 mK by a dilution refrigerator -- a substantial additional challenge compared to what is required for a 4.2 K apparatus.

\section{Apparatus Overview}

\begin{figure}[htb!]
\includegraphics[width=3in]{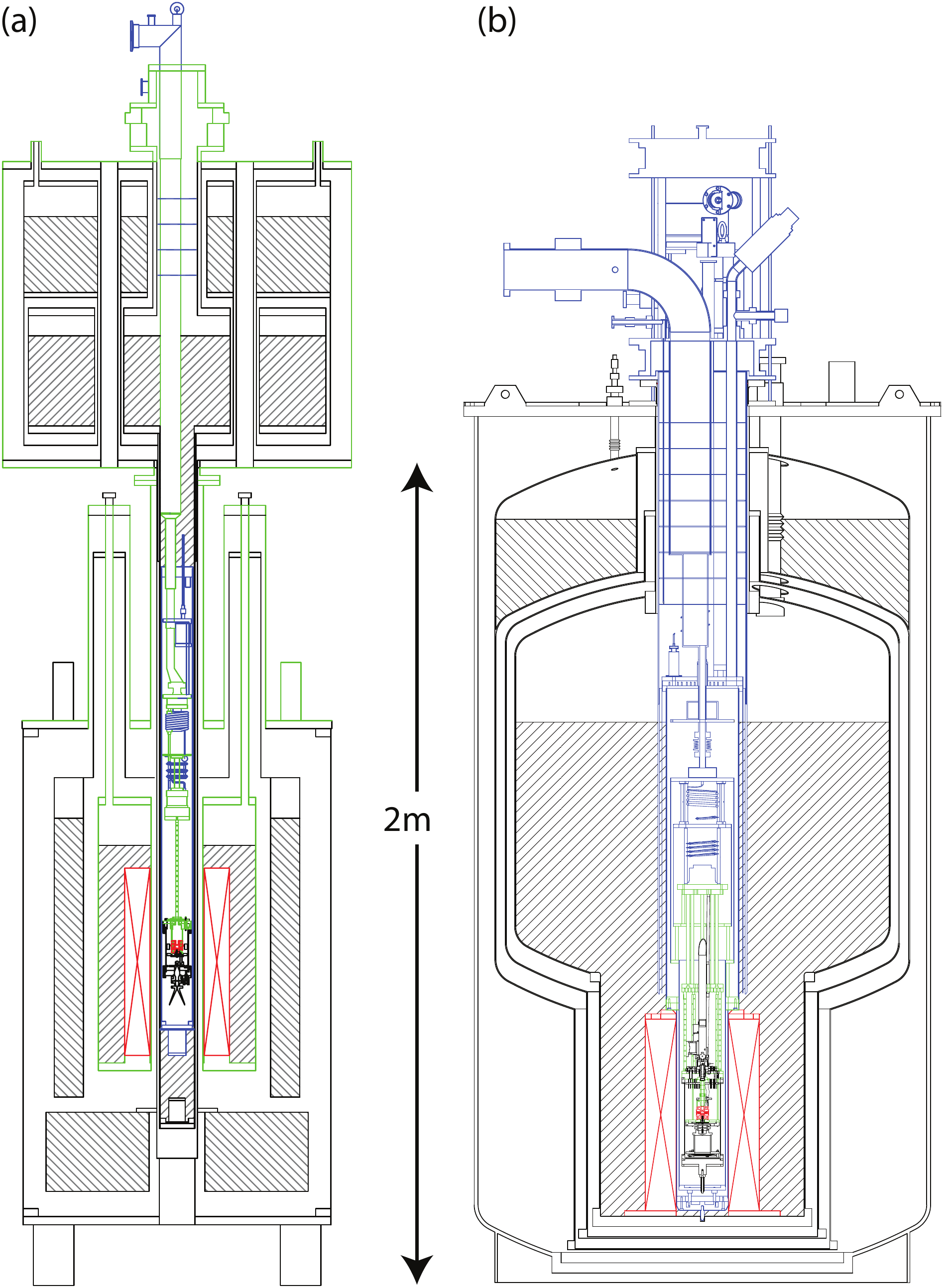}
\caption{Comparison of the old (a) and new (b) cryogenic apparatus used for comparing magnetic moments and charge-to-mass ratios of electrons and positrons. The green lines show the much shorter support path between the trap electrodes and the solenoid windings in the new apparatus compared to the old apparatus.}\label{fig:ApparatusComparison}
\end{figure}

A completely new cryogenic apparatus (Fig.~\ref{fig:WholeApparatus}) was designed for more precise measurements of the magnetic moments and charge-to-mass ratios of the positron and electron.  A custom superconducting magnet provides an access diameter of 12.8 cm (5.0 in) into which the dilution refrigerator apparatus and trap apparatus can be inserted.  This is significantly larger than the 8.6 cm (3.4 in) available diameter used for the previous measurements\cite{HarvardMagneticMoment2008,HarvardMagneticMoment2011} (illustrated in Fig.~\ref{fig:ApparatusComparison}), enough so to make it possible to admit and retract a radioactive source for positron loading. A 500 liter liquid helium reservoir is within radiation shields cooled to 77 K by a 190 liter liquid nitrogen reservoir. The persistent superconducting solenoid, sitting on the bottom of the helium reservoir and concentric with its vertical symmetry axis, produces up to a 6 tesla vertical field.

A dilution refrigerator with an attached trap is lowered into this fixed dewar from above. The so-called inner vacuum container (IVC) for this fridge is then located within the helium dewar that contains the solenoid, with its lower section located within the solenoid.  A separate and isolated trap vacuum enclosure is thermally anchored to the typically 100 mK mixing chamber of the dilution refrigerator within the IVC. 

\begin{figure}[htb!]
\includegraphics[width=2.5in]{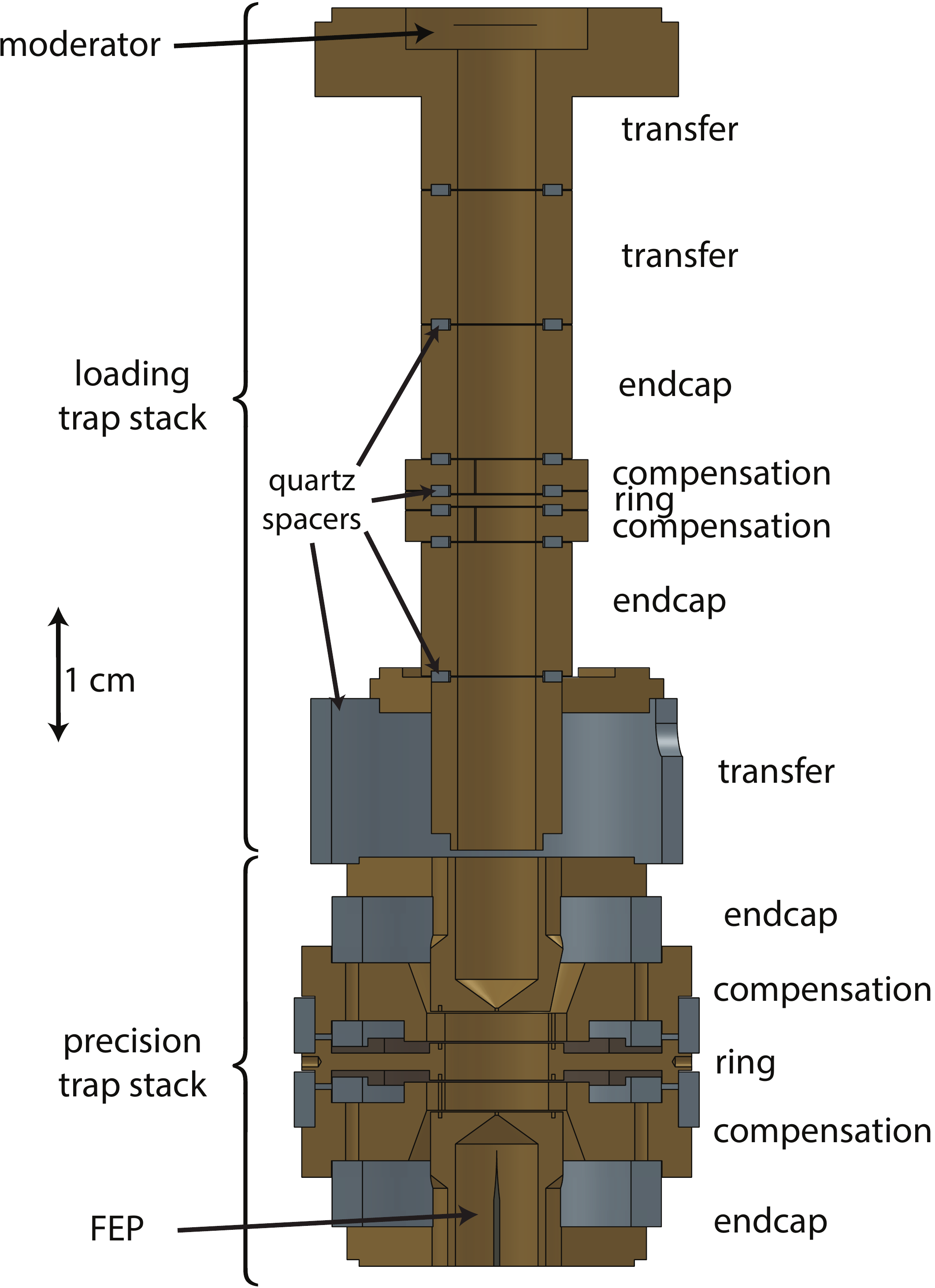}
\caption{Representation of the trap electrodes for positron loading and for precision measurements, each with an identifying label. }\label{fig:TrapDetail}
\end{figure}

The trap vacuum enclosure contains three different types of Penning traps, two of which are represented in Fig.~\ref{fig:TrapDetail}. An open-endcap cylindrical Penning trap\cite{OpenTrap} for positron loading and accumulation is at the top of the figure.  Below it is shown a closed-endcap cylindrical Penning trap\cite{CylindricalPenningTrap} for performing precision measurements with a single particle. Below the precision trap and not shown in this figure is a planar Penning trap \cite{OptimizedPlanarPenningTraps} used for scalable electron qubit studies based upon the methods developed to measure the electron and positron magnetic moments.

Compared to the apparatus we used for previous electron magnetic moment measurements, this positron-electron apparatus design has important advantages:
\begin{itemize}
\item The trap enclosure sits mechanically upon the superconducting solenoid so that the trap and magnetic field should change their mechanical position together.

\item The center axis of the trap is aligned mechanically to the center of the superconducting solenoid by pins that center the trap within the solenoid.

\item To stabilize the mechanical and electrical properties of the trap we have demonstrated the regulation of the height of the liquid helium on the trap apparatus by changing the pressure in the gas above the outer part of the liquid helium dewar \cite{ThanksDvonLindenfels}.   

\end{itemize}
However, supporting the trap from the superconducting solenoid (to keep the trap from changing its location within the superconducting solenoid) also introduce three significant cryogenic challenges.

\section{Cryogenic Challenges}

It is important that the 100 mK trap electrodes and the solenoid that produces the magnetic field for the trap be mechanically coupled so that a trapped particle and the magnetic field can only move together.  This is accomplished here by supporting the 100 mK trap directly on the 4.2 K superconducting solenoid. The trap hangs below most of the dilution refrigerator on a heat link that flexes as the trap container settles down upon the form on which the solenoid is wound.    

The first cryogenic challenge is to avoid quenching the superconducting solenoid when a warm trap apparatus is initially inserted. This is crucial insofar as the field stability required for the precise measurement is only attained if the superconducting solenoid is kept at its 4.2 K operating temperature for months.  Inserting a warm trap apparatus into the solenoid must be done so as to cool the trap and dilution refrigerator to 4.2 K before they make contact with the 4.2 K solenoid.  Otherwise the solenoid will quench, and only long after the solenoid is re-energized will the needed time stability of the field be restored. 

The trap and dilution refrigerator are slowly lowered into the large helium dewar that contains the superconducting solenoid over about 4 hours.  During this time the cold helium gas that boils off from the liquid helium cools the trap from room temperature to 4.2 K.  It is critical that no air be allowed into the dewar during the insertion.  Otherwise, the paramagnetic oxygen ice that builds up within the bore of the solenoid can both keep the trap and IVC from being fully inserted and also reduce the magnetic field homogeneity.  The procedure employs the use of a cryogenic o-ring and sliding seal, surrounded by a large plastic glove-bag.  A small continuous flow of helium gas prevents air from entering the cold volume of the apparatus.

Once the vacuum enclosure for the trap rests mechanically upon the 4.2 K solenoid, it is cooled from 4.2 K to 100 mK by turning on the dilution refrigerator.  The mechanical supports are carbon fiber posts with a very low thermal conductivity.  The conduction and radiation losses of the 100 mK vacuum container for the trap are small compared to the 330 $\mu$W that the dilution refrigerator (JDR-500 from Janis Research Company) is rated to sink at 100 mK.  

The second cryogenic challenge is that liquid helium consumption has become prohibitively expensive for a dewar of this size given the significant price increases for liquid helium in recent years. When the dilution refrigerator is operating it boils off about $19\pm 1$ liters of liquid helium per day, along with about $24\pm3$ liters of liquid nitrogen per day.  The helium consumption drops to about $9\pm 1$ liters per day when the refrigerator and trap are removed to be worked on.  The expense is considerable give that the dilution refrigerator must run for many months without stopping to make the precise magnetic moment measurements.  

Replacing the nitrogen and helium reservoirs by a pulse tube refrigerator is possible in principle but this replacement could cause significant vibration that  could affect our measurements despite the solenoid and trap being mechanically connected.  Instead we installed a helium reliquefier (Cryomech PT415 with a remote motor) that turns the cold helium gas that evaporates from the liquid helium back into liquid helium. When the helium reliquefier system is running there is essentially no net liquid helium boiled off from the apparatus.  This reliquefier does include a pulse tube refrigerator but there is some mechanical isolation from the trap and it should be possible to turn it off during the most sensitive parts of precise measurements.      


The third challenge of the cryogenic operation is preventing radiation from the warm parts of the apparatus from reaching the 100 mK trap enclosure.  The radiation must be blocked while allowing an open path for lower the radioactive source down to the 100 mK trap to load positrons, and then to retract it a distance large enough to prevent spontaneous electron loading.  To illustrate the challenge,  a 0.8 mm (1/32 inch) diameter hole that allows 300 K radiation to reach the 100 mK apparatus would provide around 200 $\mu$W of heating -- two-thirds of the total heat load that the mixing chamber of the dilution refrigerator is specified to handle.

\begin{figure}[htb!]
\includegraphics[width=2.2in]{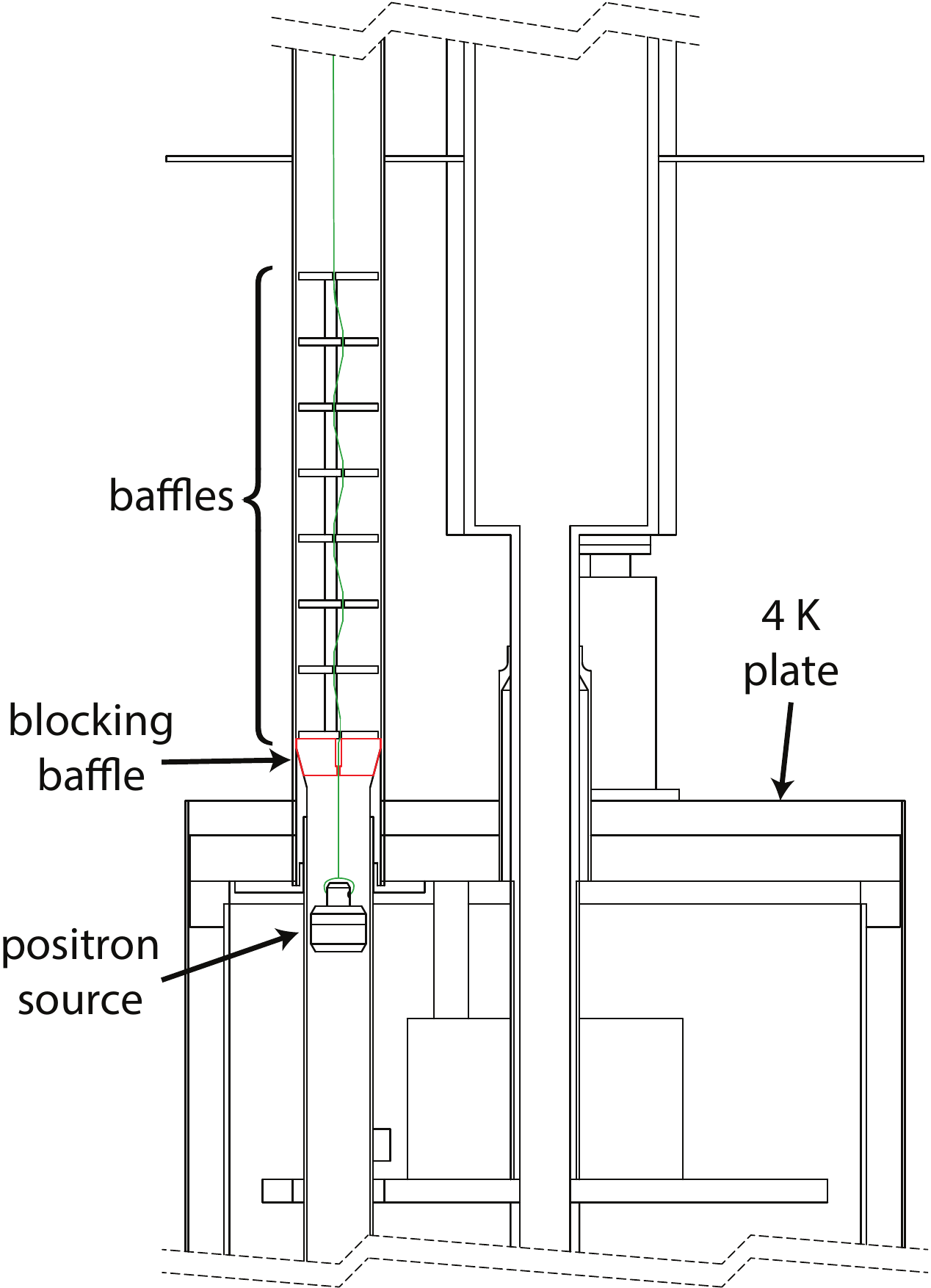}
\caption{Radiation baffles needed to avoid a thermal load on the dilution refrigerator along the entry and removal path for the radioactive source.}\label{fig:RadiationBaffles}
\end{figure}

The solution is a series of 8 baffles and a special blocking piece that float together  on the string that supports the source capsule (see Fig.~\ref{fig:RadiationBaffles}). The holes in the baffles are offset to block radiation down the center. As the source is lowered from the top of the dilution refrigerator, the blocking piece mates with a conically shape piece that is thermally anchored to the 4.2 K stage. The baffles rest on this blocking piece. This design keeps  room temperature radiation from reaching beyond the 4.2 K stage, and the 4.2 K radiation to the 100 mK trap is very small.

\section{Positron Loading}

Since the positron magnetic moment is best measured with only one trapped positron there is no need to load large numbers of positrons.  Also, positrons only need to be loaded very infrequently since once a positron is suspended within the trap for a measurement it can be used for months at a time. When positrons do need to be loaded, it is not a problem to spend an hour to accumulate tens of positrons, since only one needs to be transferred into a precision trap for measurement.  The goal of this work was thus to make sure that we could load at least one positron loaded per minute from the weakest source that could accomplish this -- preferably from a source so small that even without external shielding it could deliver a dose to an experimenter that is well below the natural background.  

\begin{figure}
\includegraphics[width=3.2in]{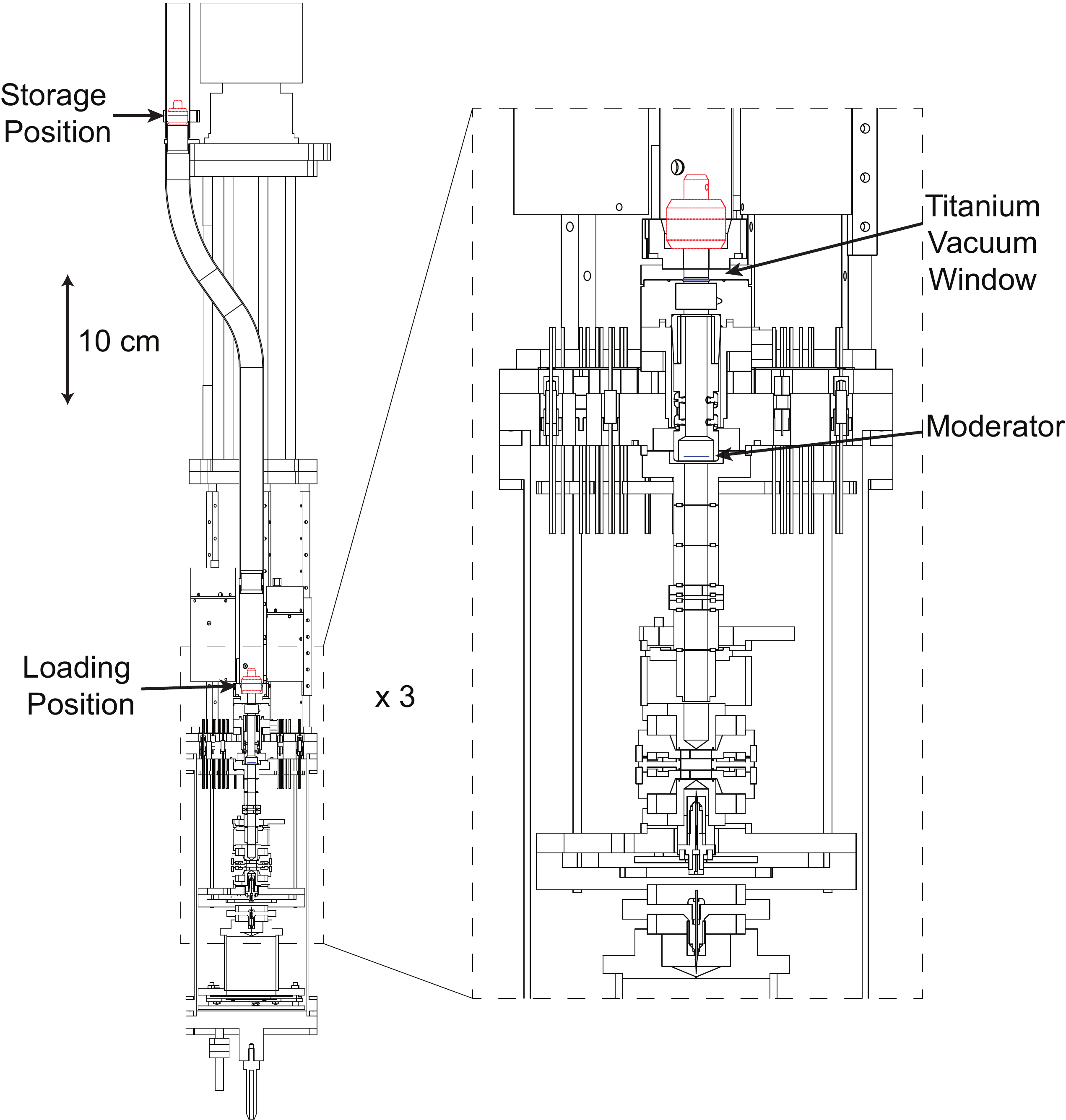}
\caption{The on-axis loading position allows positrons to be loaded into the trap. The off-axis storage position prevents positrons (or gamma-rays) from entering the trapping region.}\label{fig:sourcedeliv}
\end{figure}

We chose to use a Rydberg ionization method of accumulating trapped positrons that our group invented for its early antihydrogen experiments \cite{PositronsFromPositronium}.  Positrons thermalize within a 2 $\mu m$ thick tungsten crystal moderator located at the top of the positron loading trap as shown in Fig.~\ref{fig:sourcedeliv} and Fig.~\ref{fig:VEconfig}a. Some of the fast positrons emitted from the radioactive source enter and slow within the moderator.   These diffuse around inside the crystal until they get near a surface, whereupon the work function potential that keeps electrons within the material pops the positrons out with a low energy.  

The slow positrons that emerge follow magnetic field lines, some of them picking up an electron from the surface as they leave the thin tungsten moderator. The highly magnetized Rydberg positronium atom that is formed, only because the electron and positron are far enough apart to  mostly be pinned to field lines, is so weakly bound that it is easily ionized when the electrodes are biased to produce a strong enough electric field within the potential well of a Penning trap.  The process is quite stable even though it relies upon an adsorbed layer of gas on the exit surface for reasons that are not yet understood. By reversing the electrode potentials, this method can also be used to load electrons into the same location in the trap.   

The first of four reasons for choosing  this method is its simplicity;  positrons can be accumulated directly into high vacuum without the need for any buffer gas or any additional trapped particles. Second, the method is attractive because the accumulation rate per source activity is a couple of hundred times higher than had been realized for previous precision measurements.  Third, the method was shown to be linear in the source activity used (at least at much higher source activities than we use for this report) making it plausible that the loading rate could be extrapolated linearly to sources with less activity.  Fourth, the method has proven to be robust when the adsorbed gas is not removed from the surface. 

To estimate the source activity required to obtain the desired loading rate of about 1 e$^+$/min, we consider the loading rate achieved in earlier work in our group using a transmission moderator, which was approximately 7 e$^+$/s~\cite{PositronsFromPositronium}. The source used in this earlier work was 2.5 mCi, so the normalized loading rate was 2.8 e$^+$/s/mCi. Assuming the loading rate scales linearly with source size, we would need a 6 $\mathrm{\mu}$Ci source to achieve a 1 e$^+$/min loading rate. However, for such a small source, we would expect the self-absorption observed in the larger source \cite{HaarsmaThesis} to be minimal and thus would expect the loading rate to be increased by a factor of 2 from the earlier work. Therefore we estimate a 3 $\mathrm{\mu}$Ci source will give a 1 e$^+$/min loading rate, and equivalently, a 6 $\mathrm{\mu}$Ci source will give a 2 e$^+$/min loading rate. 

\begin{figure}[htb!]
\includegraphics[width=1.5
in]{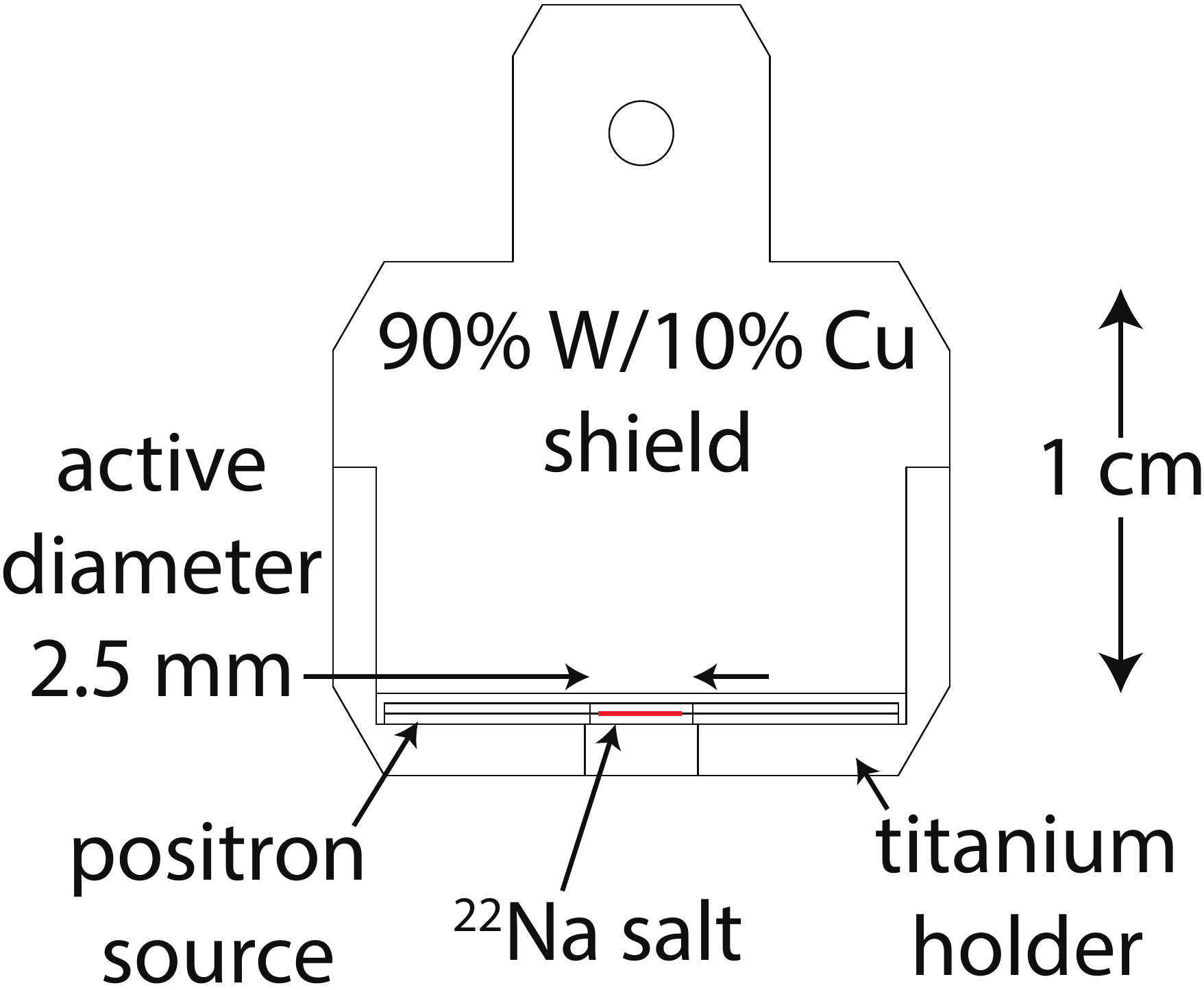}
\caption{The sealed button source is inside the two-piece capsule used to provide shielding and allow manipulation of the source.}\label{fig:SourceCapsule}
\end{figure}

The positron source we use in this work is a $^{22}$Na sealed button source (Isotope Products Laboratories custom diameter POSN source).  Its activity was 15.6 $\mathrm{\mu}$Ci when delivered, and it decreased between 6.9 and 6.3 $\mathrm{\mu}$Ci during the studies reported here.  The delivered source had the radioactive salt between two 5 $\mu m$ thick Ti foils that were electron-beam welded to completely contain the salt.  To facilitate safe handling, we enclosed this source in a two-piece capsule made from a 90\% tungsten/10\% copper composite, as represented in Fig.~\ref{fig:SourceCapsule}. 

A nylon string connects the top source capsule to a rotational vacuum feedthrough at the top of the dilution refrigerator. This allows the source to be raised and lowered inside the inner vacuum chamber of the dilution refrigerator. Pairs of light-emitting diodes and photo-diodes, as well as marks on the string, are used to monitor the location of the source. The positron source can be raised and lowered along the entire length of the dilution refrigerator, although in normal operation it is moved between two locations: an off-axis ``storage" position at the mixing chamber of the dilution refrigerator where positrons are prevented from loading into the trap by the distance and the angle; and an on-axis ``loading" position where the source capsule is located directly above the trap vacuum enclosure, and separated from it by a thin titanium foil vacuum window. These locations are shown in Fig.~\ref{fig:sourcedeliv}.

\section{Results and Discussion}

Particles in the loading trap are detected and counted nondestructively using a cryogenic amplifier connected to a trap electrode. The amplifier and trap electrodes form an RLC circuit and we monitor the Johnson noise spectrum of this circuit. When the trap is empty, the frequency spectrum is a Lorentzian peak centered at the circuit's resonant frequency. The oscillation of a small number of particles in the trap interacts with the tuned circuit resonance and forms a Lorentzian ``dip'' in the amplifier resonance whose center is the particles' axial oscillation frequency and whose FWHM is equal to the number of particles in the trap multiplied by the single-particle damping width, $\gamma_z$. 

\begin{figure}
\includegraphics[width=3in]{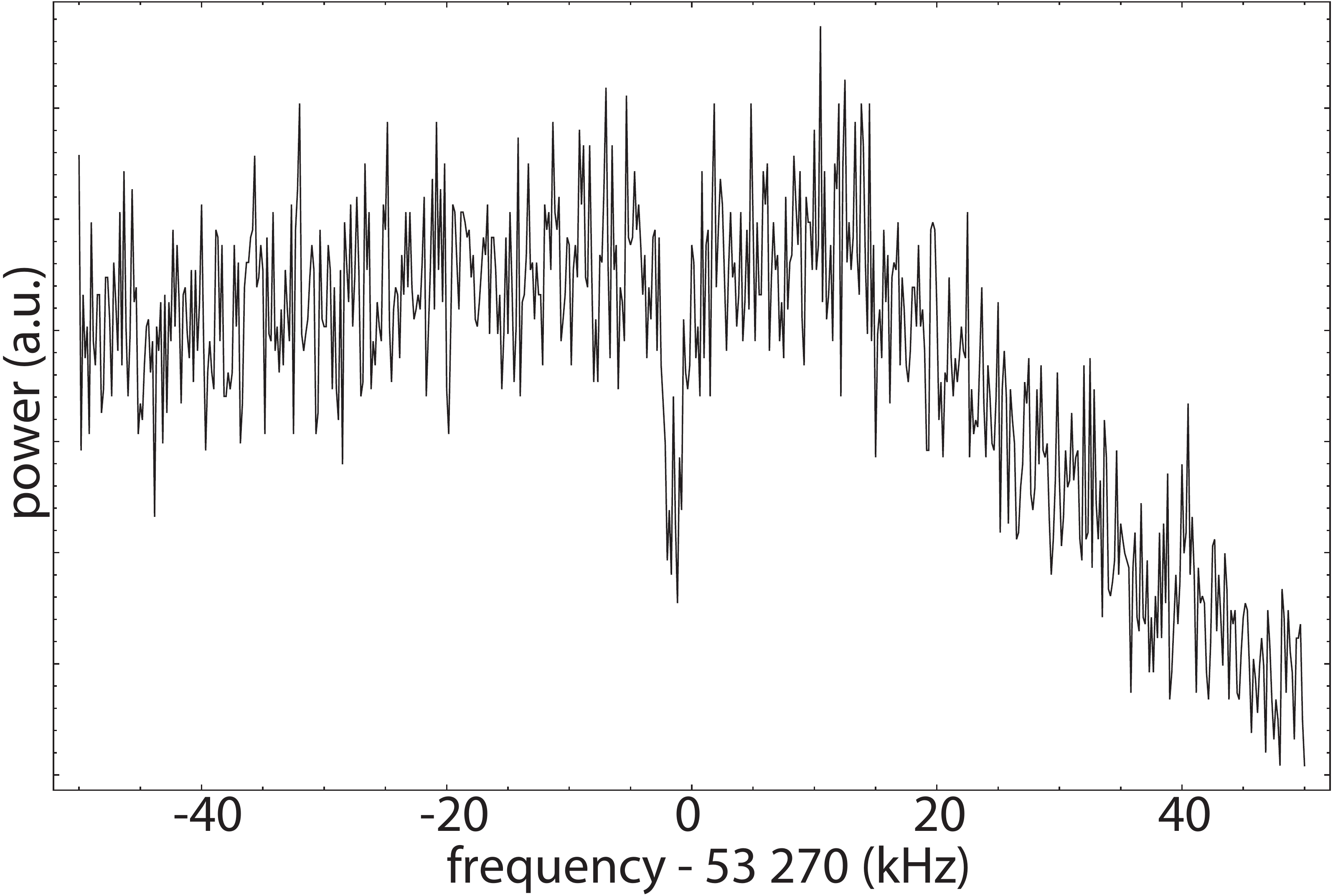}
\caption{An example of the dip in the amplifier noise resonance for a cloud of approximately 160-320 positrons. }\label{fig:dip}
\end{figure}

We have calculated that $\gamma_z = 10$ Hz for this amplifier. The actual value of $\gamma_z$ tends to be somewhat smaller than the calculated value\cite{Review} and we have some experimental indications that $\gamma_z \approx 5$ Hz. Therefore we use the estimate of $\gamma_z =$ 5-10 Hz in discussing the number of positrons (or electrons) in this paper. Figure \ref{fig:dip} shows a cloud of approximately 160-320 positrons. It should eventually be possible to remove the large uncertainty in the number of positrons in the accumulation trap by optimizing the detection amplifier so that it unambiguously resolves the difference between one and two trapped positrons.

\begin{figure}[htbp!]
\includegraphics[width=3in]{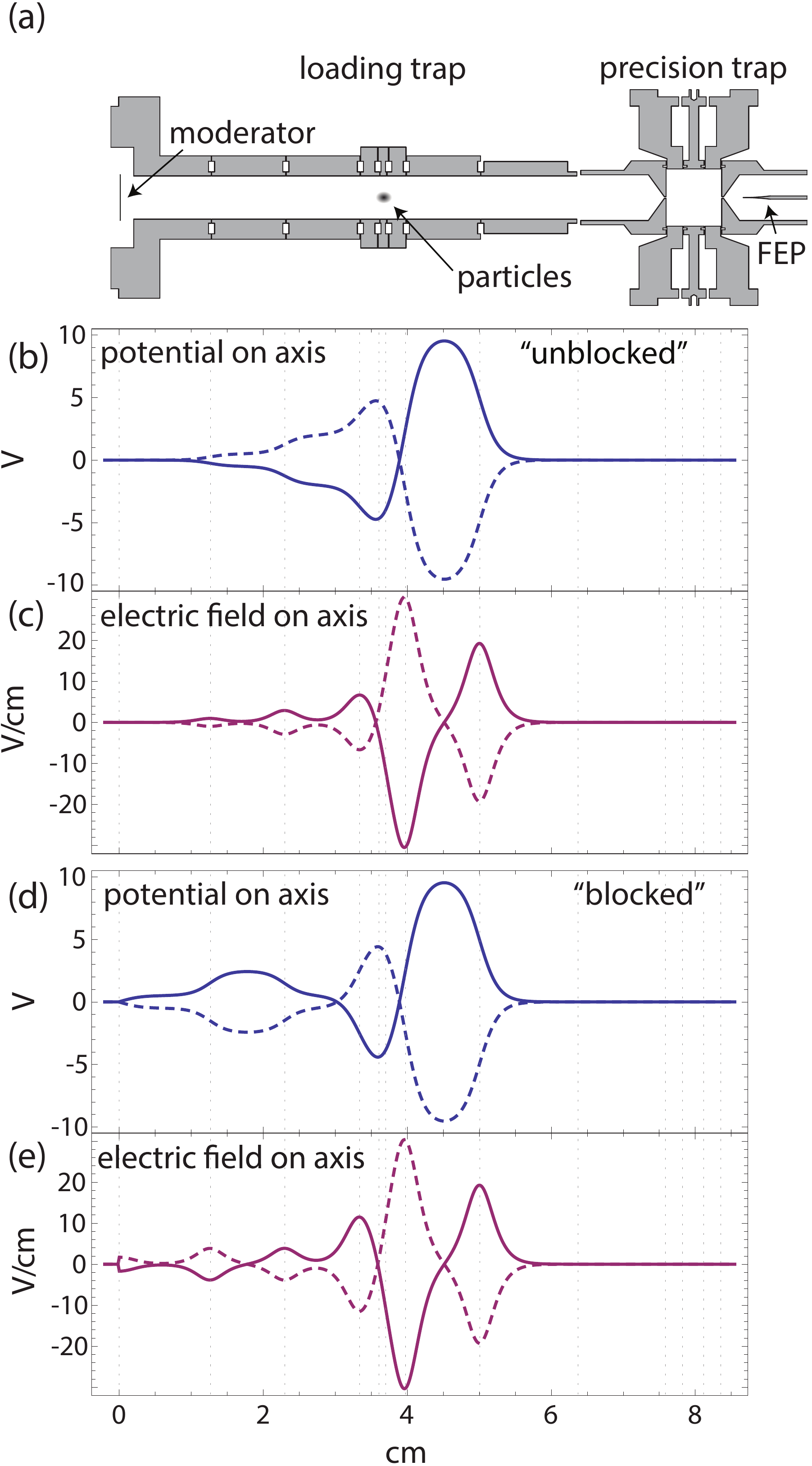}
\caption{The voltage and electric field profiles are shown for the two main loading configurations used. Solid lines denote the potential and field used for positron loading and dashed lines denote the potential and field used for electron loading. Figures (b) and (c) show the configuration where secondary electrons are unblocked (not prevented from entering the trapping region) while (d) and (e) show the configuration where secondary electrons are blocked (prevented from entering the trapping region). A cross section of the electrode stack is shown (top).}\label{fig:VEconfig}
\end{figure}

Both positrons and electrons have been loaded from our positron source, and the numbers loaded have been used to verify the loading mechanism. Several different electrode potential configurations were used for these loading trials. The main potential configurations are shown in Fig.~\ref{fig:VEconfig}. 

\begin{figure}
\centering
\includegraphics[width=3.3in]{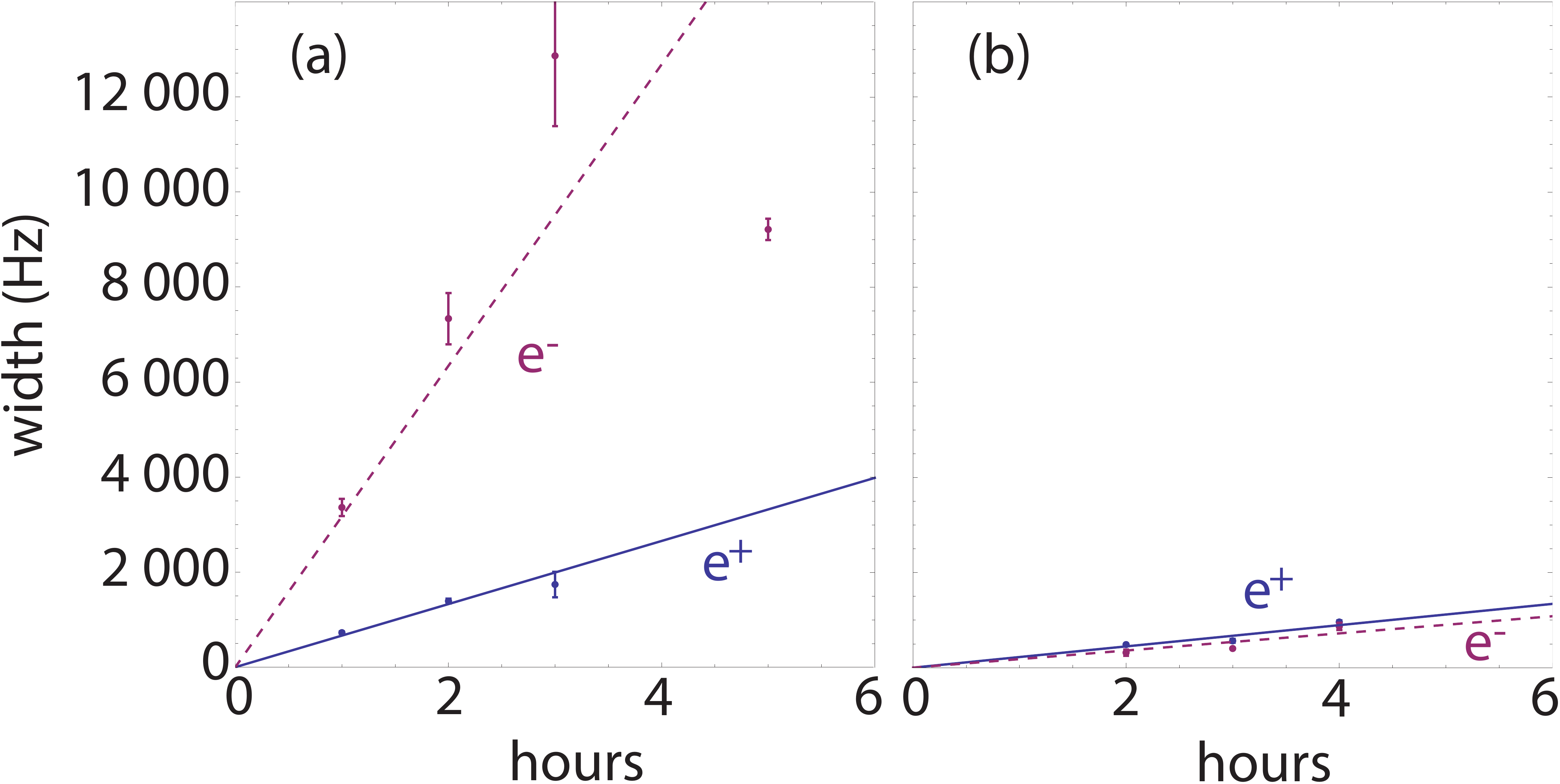}
\caption{Electron (dashed lines) and positron (solid lines) loading rates for the two electrode potential configurations shown in figure \ref{fig:VEconfig}. Figure (a) shows the configuration where secondary electrons are unblocked and (b) shows the configuration where secondary electrons are blocked. When secondary electrons are prevented from entering the trapping region, (b), the large asymmetry in electron and positron loading rates, (a), is greatly reduced.}\label{fig:eploadingrates}
\end{figure}

For the field ionization method of accumulating positrons or electrons from positronium, the loading rate for positrons and electrons from the source into the trap should be identical.  This is provided that the electrodes are biased to prevent loading the trap with secondary electrons knocked free of the moderator in collisions with positrons passing through. Additionally, the loading rate should be linear in time as the loading mechanism does not depend upon interactions with previously loaded particles. In all cases, the number of particles we observed was indeed linear in time. We found that changing the electrode potentials from the ``unblocked'' configuration shown in figure \ref{fig:VEconfig}, which does not prevent secondary electron loading, to the ``blocked'' configuration, which prevents most secondary electron loading, greatly reduces the large asymmetry seen in electron and positron loading rates as seen in Fig.~\ref{fig:eploadingrates}.  This is as expected for the loading mechanism discussed.  

Additional factors also affect the loading rate as expected. For example, moving the source 3.8 cm (1.5 in) further from the moderator and loading trap cuts the loading rate of both electrons and positrons in half. Additionally, firing a field emission point (or FEP, located at the bottom of the precision trap and used for quickly loading electrons into either trap) 20 times at approximately 1 nA for approximately  1 minute each time decreased the loading rate by roughly a factor of two. This is because the electron beam from the field emission point strikes the surface of the moderator, and the formation of positronium at the moderator surface depends upon a layer of adsorbed gas on the moderator surface. This phenomenon was observed previously when antiprotons struck such a moderator\cite{ThesisBowden} and when a laser was used to deliberately heat the moderator\cite{ThesisEstrada}. Thermal cycling of the apparatus restores the adsorbed gas layer and restores the loading rate.

Our maximum positron loading rate manifests itself as a 687 (10) Hz/hr increase in the measured resonance width, corresponding to about 70-140 e$^+$/hr being accumulated in the trap. For our 6.3 $\mathrm{\mu}$Ci $^{22}$Na source, this gives a loading rate of 3-6 e$^+$/s/mCi, in good agreement with previous work\cite{PositronsFromPositronium}. 

\section{Conclusion}

Field ionization of slow positronium accumulates the positrons needed to measure the positron magnetic moment at the $3$ parts in $10^{13}$ precision to which we have measured the electron magnetic moment.  Compared to the positron accumulation used for previous magnetic moment measurements, an orders of magnitude smaller source accumulates trapped positrons at a rate per source activity that is orders of magnitude larger.  The tiny 6.3 $\mu$Ci source that provides positrons to a 100 mK Penning trap is small enough to avoid special licensing and handling methods, provides a dose rate to an unshielded user that is much lower than the average natural background, and can be withdrawn from the trap during precise measurements.     

\begin{acknowledgments}
This work was supported by the NSF.  
\end{acknowledgments}


%

\end{document}